\documentclass[11pt,a4paper]{article}
\usepackage{jheppub}

\usepackage[dvipsnames]{xcolor}
 \usepackage{tikz}

\usepackage{graphicx,verbatim}
\usepackage{amsmath}
\usepackage{amsfonts}
\usepackage{amssymb}
\usepackage{psfrag}
\usepackage{accents}
\usepackage[T1]{fontenc}

\usepackage[colorinlistoftodos]{todonotes}
\usepackage[normalem]{ulem}
\usepackage{epstopdf}

\newcommand{\pb}[1]{\hbox{\lower0.5ex\hbox{${}_{\leftarrow}$}}\kern-1.9ex{#1}}

\def\={\hat{=}}

\def\be{\begin{equation}}
\def\ee{\end{equation}}
\def\ba{\begin{eqnarray}}
\def\ea{\end{eqnarray}}

\def\SO(3){\rm SO(3)}
\def\so(3){\rm so(3)}
\def\SO(4){\rm SO(4)}
\def\so(4){\rm so(4)}
\def\SO(1,4){\rm SO(1,4)}
\def\so(1,4){\rm so(1,4)}
\def\SU(2){\rm SU(2)}

\usepackage{amsthm}
\theoremstyle{remark}
\newtheorem{remark}{Theorem}
\newtheorem{theorem}{Theorem}[section]

\newtheorem{lemma}[theorem]{Lemma}

\begin{document}

\title{Towards the black hole uniqueness: transverse deformations of the extremal Reissner--Nordstr\"{o}m--(A)dS horizon}
%{Quasi-local Horizons: Universal Structures, Symmetry Groups and Associated Fluxes} 
\author{Maciej Kolanowski}
\affiliation{Institute of Theoretical Physics, Faculty of
  Physics, University of Warsaw, Pasteura 5, 02-093 Warsaw, Poland}
  \emailAdd{Maciej.Kolanowski@fuw.edu.pl}
\abstract{ \noindent
We study all transverse deformations of the extremal Reissner--Nordstr\"{o}m--(A)dS horizon in the Einstein--Maxwell theory. No symmetry assumptions are needed. It is shown, that for the generic values of a charge, the only allowed deformation is spherically symmetric. However, it is shown that for fine-tuned values of the charge, the space of deformations is larger, yet still finite-dimensional.
}
%\pacs{04.70.Bw, 04.25.dg, 04.20.Cv}
%\pacs{98.80.Qc, 04.60.Pp, 04.60.Kz}
\maketitle
\section{Introduction} \noindent
The idea of the black hole uniqueness started with the seminal work of Israel \cite{PhysRev.164.1776}. The topic was heavily investigated for over fifty years now with a notable results from Hawking \cite{hawking1972black}, Carter \cite{Carter:1971zc}, Robinson \cite{PhysRevLett.34.905}, Mazur \cite{Mazur1982}, Chruściel and Wald \cite{Chrusciel1994}. For the modern review, see \cite{chrusciel2012stationary}. \\
Killing horizons, which are central objects in virtually all uniqueness theorems, can be divided into two categories: non-degenerate or degenerate (extremal) ones, on which Killing vector fields are affinely parametrized. This second type required separate analysis, in particular solving the infamous Near Horizon Geometry equation \eqref{nhg}. In this paper we will focus entirely on the degenerate case, albeit from a different perspective: given a solution to \eqref{nhg}, in what ways can we extend it to the bulk? \\
The question of the black holes' uniqueness changes drastically when one introduces a cosmological constant. Indeed, there is a plethora of static solutions to the Einstein-Maxwell-$\Lambda$ equations which are asymptotically AdS, and they do not possess any spatial symmetry \cite{Herdeiro:2016plq}.  On the other hand, it may be shown that the only static NHG with the electromagnetic field is given by the extremal Reissner-Nordstr\"om-(A)dS \cite{Kunduri:2008tk}, thus suggesting that the extremal case is more rigid when $\Lambda \neq 0$. We will further strengthen this result, showing that the first-order deformations must be spherically-symmetric as well.
\subsection{Near Horizon Geometry limit} \noindent
On the neighborhood of the extremal horizon, one can introduce a null gaussian coordinate system in which the metric reads:
\begin{equation}
    g = 2dv \left( 
    dr + rh_a dx^a + \frac{1}{2}r^2 F dv
    \right) + \gamma_{ab}dx^a dx^b,
\end{equation}
where the horizon is located at $r=0$, $x^a$ are coordinates on the cross-section (a sphere, in our case) and $F, h_a, \gamma_{ab}$ depend upon $r$ and $x^a$ only. The Killing vector field is given by $\partial_v$. Maxwell field around $r=0$ admits a similar form\footnote{Einstein equations implies that $\mathcal{F}_{va}$ pull-backed to the horizon vanishes}:
\begin{equation}
    \mathcal{F} = \Psi dv \wedge dr + rW_a dv \wedge dx^a + Z_a dr \wedge dx^a + \frac{1}{2} B_{ab} dx^a \wedge dx^b,
\end{equation}
where $\Psi, W_a, Z_a, B_{ab}$ are smooth and depend only upon $r, x^a$. \\
We may now consider a one-parameter family of diffeomorphisms 
\begin{equation}
    \phi_\epsilon (v,r, x^a) = (\epsilon^{-1}v, \epsilon r, x^a)
\end{equation}
for $\epsilon > 0$. Quite remarkably, objects $g(\epsilon):=\phi_\epsilon^\star g$ and $\mathcal{F}(\epsilon):=\phi_\epsilon^\star \mathcal{F}$ possess well-defined limit when $\epsilon \to 0$. Since $\phi_\epsilon$ is no longer diffeomorphism in this limit, it may be an entirely new (yet much simpler) solution to the Einstein-Maxwell equations. Indeed, we can write those limits as
\begin{align}
\begin{split}
    \lim_{\epsilon\to 0} \phi^\star_\epsilon g &= 2dv \left( 
    dr + rh_a^{(0)} dx^a + \frac{1}{2}r^2 F^{(0)} dv
    \right) + \gamma_{ab}^{(0)}dx^a dx^b \\
    \lim_{\epsilon\to 0} \phi^\star_\epsilon \mathcal{F} &= \Psi^{(0)}dv \wedge dr + rW_a^{(0)} dv \wedge dx^a + \frac{1}{2} B_{ab}^{(0)} dx^a \wedge dx^b,
\end{split}
\end{align}
where all objects $h_a^{(0)}, F^{(0)}, \Psi^{(0)}, W_a^{(0)}, B_{ab}^{(0)}$ are now $r$ independent, and they are given by the limits as $r\to 0$ of associated objects in the original spacetime. Einstein-Maxwell equations in this setting read
\begin{align}
    &R_{ab}^{(0)} = \frac{1}{2} h_a^{(0)} h_b^{(0)} - D_{(a} h_b^{(0)} + \Lambda \gamma_{ab}^{(0)} + 2 B_{ac} B_b^{(0)\ c} + \frac{2}{D-2} \gamma_{ab}^{(0)} \Psi^{(0) 2} - \frac{1}{D-2} \gamma_{ab}^{(0)} B^{(0) 2}  \label{nhg} \\
    &F^{(0)} = \frac{1}{2} h^{(0)\ 2} - \frac{1}{2}D^a h_{a}^{(0)} + \Lambda - 2\frac{D-3}{D-2} \Psi^{(0)2} - \frac{1}{D-2}B^{(0)2} \\
    &W^{(0)} = d\Psi^{(0)} \\
    &dB^{(0)} = 0
    \\
    &\left(
    D_a - h_a^{(0)}
    \right)\Psi^{(0)} + \left(
    D^b - h^{b(0)}
    \right) B_{ba} = 0,
\end{align}
where $D$ is the number of spacetime dimensions, $D_a$ is the covariant derivative associated with $\gamma^{(0)}$ and all indices are lowered and raised using $\gamma^{(0)}$. Those equations, although easier than the full set of Einstein-Maxwell equations, are still being studied, especially in 4 and 5 dimensions. Nevertheless, significant results were achieved. For example, in four dimensions they allowed to classify all possible topologies of the horizons in the vacuum case \cite{Dobkowski-Rylko:2018nti}, also all axially-symmetric solutions were found \cite{Lewandowski:2002ua}. In higher dimensions (first considered in \cite{Lewandowski:2004sh}) much less is known because those equations are much less restrictive. For a recent review, see \cite{Kunduri:2013gce}.
\subsection{Deformations} \noindent
In \cite{Li:2015wsa} it was proposed to systematically study extremal black holes in the small $\epsilon$ limit and equations for the first order terms were derived. They were generalized to include matter fields in \cite{Fontanella:2016lzo, Li:2018knr}. They can be obtained from the Einstein-Maxwell equations satisfied by $(g(\epsilon), \mathcal{\epsilon})$ and differentiating them with respect to $\epsilon$. Those equations, although linear, are rather lengthy and not especially illuminating. Their special case is presented below as \eqref{einstein} and \eqref{maxwell}. Interested readers should consult original papers on the topic. \\
One could naively think that those results are valid only in linearized gravity, since they involve expansion in the small parameter. However, these equations also appear as part of the exact constraints on the characteristic Cauchy data defined on a degenerate (extremal) Killing horizon [8], (or, more generally, on an extremal isolated horizon that is Killing to the appropriate order). The knowledge of this connection is necessary to obtain a well-posed characteristic initial value problem. On the other hand, one could truly interpret solutions to \eqref{einstein} and \eqref{maxwell} as zero-modes living on the background of the near-horizon geometry. Thus, they could be of some interest in the holographic context. \\
The rest of the paper is organized as follows: in Sec. \ref{class} we present the most general deformation and sketch reasoning behind the derivation. We discuss its properties in Sec. \ref{discussion}. Technical details are relegated to the Appendix \ref{tech_app}.
\section{Classification of the transverse deformations} \label{class}
\subsection{Background} \noindent
We consider the following solution to the Einstein--Maxwell equations\footnote{Notice that for simplicity we changed notation with respect to the previous Section}:
\begin{align}
    \begin{split}
        g &=2dv \left( dr + \frac{1}{2} r^2 F dv
        \right) + r^2_{+} \left(
        \frac{dx^2}{1-x^2} + (1-x^2) d\phi^2
        \right)\\
        \mathcal{F} &= \Psi dv \wedge dr, \label{back}
    \end{split}
\end{align}
where $\Psi = -\frac{Q}{r_+^2}$, $Q$ is an electric charge, $F=\Lambda - \Psi^2$ and $r_+$ is an area radius of the horizon. They satisfy
\begin{equation}
    \frac{1}{r^2_+} = \Lambda +\Psi^2 
\end{equation}
and so
\begin{equation}
    r_+ = \sqrt{\frac{2Q^2}{1 \pm \sqrt{1 - 4Q^2 \Lambda}}}.
\end{equation}
Solution \eqref{back} can be obtained as a near-horizon limit of the extremal Reissner--Nordstr\"om-(A)dS black hole. It will serve a r\^ole of the background for the horizon deformations (which can be seen as well as a zero-modes in this spacetime). For future convenience, let us denote metric induced on the surfaces $r,v =\textrm{const.}$ as $q = r_+^2 \mathring{q}$. We will denote geometrical objects associated with $\mathring{q}$ by a small circle. \\
A few remarks are in place:
\begin{itemize}
    \item[(i)] It is clear that when $\Lambda > 0$, there is an upper bound on the charge. It should not come as a surprise, since the mass is bounded as well.
    \item[(ii)] When $\Lambda > 0$, for the given charge, there are two possible values of $r_+$. Indeed, generically, a Reissner--Nordstrom-dS black hole posses three horizons. Extremal case corresponds to the merger of either two of those.
    \item[(iii)] Solution \eqref{back} is not the most general spherically symmetric near horizon geometry in the Einstein--Maxwell theory, since one could employ magnetic field as well. The general case reads:
    \begin{align}
    \begin{split}
        g &=2dv \left( dr + r^2 (\Lambda - \Psi^2 - \frac{1}{2} B_{ab}B^{ab}) dv)
        \right) + r^2_{+} \left(
        \frac{dx^2}{1-x^2} + (1-x^2) d\phi^2
        \right)\\
        \mathcal{F} &= \Psi dv \wedge dr + \frac{1}{2}B_{ab} dx^a \wedge dx^b, \label{general}
    \end{split}
\end{align}
where $B_{ab} = Q_m \mathring{\epsilon}_{ab}$ and the radius is given by
    \begin{equation}
    \frac{1}{r^2_+} = \Lambda +\Psi^2 + \frac{1}{2}B_{ab} B^{ab}.
\end{equation}
However, the latter can be obtained from the former by a simple dual symmetry:
\begin{align}
    \begin{split}
        Q &\mapsto \sqrt{Q_e^2 + Q_m^2} \\
        \mathcal{F} &\mapsto \frac{q_e}{\sqrt{Q_e^2 + Q_m^2}} \mathcal{F} - \frac{Q_m}{\sqrt{Q_e^2 + Q_m^2}} \star \mathcal{F},
    \end{split}
\end{align}
where $\star$ is a Hodge dual associated with $g$.
Since this field rotation is a symmetry of Einstein--Maxwell theory, we can assume $B_{ab} = 0$ without loss of generality. Indeed, one can simply apply dual transformation to the deformed solution, which would transform background and deformation separately.  
\item[(iv)] In higher dimensions, the aforementioned magnetic term breaks the spherical symmetry, since there is no spherically symmetric two-form $B_{ab}$ on $\mathbb{S}^d$ when $d>2$. Thus, our choice shall allow for a simpler generalization to the higher-dimensional horizons. We will discuss in more details in the Sec. \ref{discussion}.
\end{itemize}
\subsection{Deformations} \noindent
We can now consider transverse deformations of \eqref{back}. Our discussion will follow closely the derivation of \cite{Li:2018knr} with a small changes in notation. At the leading order, the general form of the deformation is
\begin{align}
    \begin{split}
        \delta g &= r^3 F^{(1)} dv^2 + 2 r^2 h_{a}^{(1)} dv dx^a + r \gamma_{ab}^{(1)} dx^a dx^b \\
        \delta \mathcal{F} &= r \Psi^{(1)} dv \wedge dr + r^2 W_a^{(1)} dv \wedge dx^a + Z_a^{(1)} dr \wedge dx^a + \frac{1}{2}r B_{ab}^{(1)} dx^a \wedge dx^b.
    \end{split}
\end{align}
All terms written above are $r$ and $v$ independent. Data $(F^{(1)}, h_a^{(1)}, \Psi^{(1)}, W_a^{(1)}, B_{ab}^{(1)})$ can be obtained algebraically from $\gamma^{(1)}_{ab}, Z^{(1)}_a$ as follows:
\begin{align}
    \begin{split}
        h_{a}^{(1)} &= - \frac{1}{2}D^b \gamma_{ab} + \frac{1}{2} D_a \gamma + 2 \Psi Z_a^{(1)} \\
        F^{(1)} &=-\frac{1}{3}D^a h^{(1)}_a - \frac{1}{6} (\Lambda - \Psi^2) \gamma - \frac{2}{3} \Psi \Psi^{(1)} \\
        \Psi^{(1)} &= D^a Z_a - \frac{1}{2} \Psi \gamma^{(1)} \\
        W^{(1)} &= \frac{1}{2} d \Psi^{(1)} \\
        B^{(1)} &= dZ^{(1)},
    \end{split}
\end{align}
where $\gamma^{(1)} = q^{ab} \gamma_{ab}^{(1)}$.
From the Einstein equations, the remaining equation for $\gamma_{ab}^{(1)}$ follows\footnote{Notice that those equations are already evaluated at the background of \eqref{back} and thus are much simpler.}
\begin{align}
    \begin{split}
        0 &= \Delta_L \gamma_{ab}^{(1)} + \frac{1}{2} D_{(a}D_{b)}\gamma^{(1)} - 2\Psi^2 \left( \gamma_{ab}^{(1)} - \frac{1}{2} \gamma^{(1)} q_{ab} \right) + 4\Psi \left(
        D_{(a}Z_{b)} - \frac{1}{2} D^c Z_c q_{ab}
        \right), \label{einstein}
    \end{split}
\end{align}
where $\Delta_L \gamma^{(1)}_{ab} = -\frac{1}{2} \Delta \gamma_{ab} + R \left(\gamma_{ab}^{(1)} - \frac{1}{2}\gamma^{(1)} q_{ab} \right)$ is a Lichnerowicz operator associated with $q_{ab}$ and $\Delta = q^{ab} D_a D_b$. Notice that these equations are automatically traceless. Maxwell equations on $Z_a$ on the other hand read:
\begin{equation}
    0 = \Delta Z_a^{(1)} - \frac{1}{r^2_{+}} Z_a^{(1)} - 4 \Psi^2 Z_a^{(1)} - 2F Z_a^{(1)} + \Psi D^b \gamma_{ab}^{(1)} - \frac{3}{2} \Psi D_a \gamma^{(1)}. \label{maxwell}
\end{equation}
Our theory can be regarded as a linearized gravity, and thus it enjoys a large group of gauge transformations:
\begin{align}
    \begin{split}
        \delta g \mapsto \delta g + \mathcal{L}_\xi g \\
        \delta \mathcal{F} + \mathcal{L}_\xi \mathcal{F}.
    \end{split}
\end{align}
Since we consider a very specific form of $\delta g, \delta \mathcal{F}$, vector fields $\xi$ generating gauge transformations must preserve it. It is quite a severe restriction, the most general allowed $\xi$ (modulo isometries of $f$) is:
\begin{equation}
    \xi = -\frac{1}{2}f\partial_v + \frac{1}{2} r D^a f \partial_a,
\end{equation}
where $f = f(x^a)$ is an arbitrary smooth function on a sphere. Geometrically, gauge symmetry is equivalent to the freedom in choosing a cross-section $v=0$ of the horizon. Action of $\xi$ can be explicitly written as
\begin{align}
    \begin{split}
        \gamma_{ab}^{(1)} &\mapsto \gamma_{ab}^{(1)} + D_a D_b f \\
        h_a^{(1)} &\mapsto h_a^{(1)} - \frac{1}{2}F D_a f \\
        F^{(1)} &\mapsto F^{(1)} + \frac{1}{2} + \frac{1}{2} D^a f D_a F \\
        \Psi^{(1)} &\mapsto \Psi^{(1)}\\
        Z_a^{(1)} &\mapsto Z_a^{(1)} + \frac{1}{2} \Psi D_a f.
    \end{split}
\end{align}
It is easy to check that the equations of motion are invariant with respect  to those transformations. Thus, it may be concluded that Eq. \eqref{einstein} and \eqref{maxwell} are undetermined. However, it was shown in \cite{Fontanella:2016lzo} that they become elliptic once properly gauged (see \cite{Li:2015wsa} for a proof in the absence of an electromagnetic field). It follows from the Fredholm theory that the moduli of solutions (up to the gauge transformations) is finite dimensional. One globally available gauge condition is $\gamma^{(1)} = \textrm{const.}$ However, we will not impose it but rather work with gauge invariant potentials for $(\gamma_{ab}^{(1)}, Z^{(1)}_a)$ from the start. 
\subsection{Solutions} \noindent
In this section we will write down the solutions to Eq. \eqref{einstein} and \eqref{maxwell} and discuss when they exist. Derivation, which is quite lengthy, is redirected to the Appendix \ref{tech_app}. \\
The most general solution to the Eq. \eqref{maxwell} is:
\begin{align}
    \begin{split}
        \gamma_{ab}^{(1)} &= -2\left(
        \mathring{\epsilon}_{c(a}\mathring D_{b)} \mathring D^c
        \right) \left(\mathring{\Delta} - 2 \right) \chi + \left(
        \mathring{D}_a \mathring{D}_b - \frac{1}{4} q_{ab} (\mathring\Delta -2)
        \right) \Phi +\mathring{D}_a\mathring D_b f\\
        Z_{a} &= -\Psi \epsilon_{ac} \mathring D^c (\mathring\Delta +2) \chi + \frac{1}{2}\Psi \mathring{D}_a f,
    \end{split}
    \label{sol_to_max}
\end{align}
where $\mathring{D}$ is a covariant derivative associated with $\mathring{q}$, $\mathring\Delta = \mathring{q}^{ab} \mathring D_a \mathring D_b$ and $\chi$ does not have $l=1$ components in the decomposition on the spherical harmonics\footnote{from now by $l$ — components we will mean components in the aforementioned decomposition}, it means
\begin{equation}
    \int_{S^2} Y_{1m}^\star \chi = 0.
\end{equation}
Moreover, notice that $l=1$ component of $\Phi$ and $l=0$ component of $\chi$ do not affect $(\gamma^{(1)}_{ab}, Z^{(1)}_a)$ and thus we will set them to zero for simplicity. Inserting those solutions to \eqref{einstein} leads to
\begin{align}
\begin{split}
    \left(\mathring\Delta -2 + 8 r_{+}^2 \Psi^2 \right) \Phi &= \textrm{const.} \\
    \mathring\Delta \left(
    \mathring\Delta -2 + 8 r_{+}^2 \Psi^2
    \right) \chi &= 0
    \label{final_eq}
\end{split}
\end{align}
Thus, there is always a $\Phi = \textrm{const.}$ solution. For generic $Q, \Lambda$ operator $\left(
    \mathring\Delta -2 + 8 r_{+}^2 \Psi^2
    \right)$ is invertible, and thus it is the only solution. It is not invertible only when
    \begin{equation}
        4 Q^2 \Lambda = 1 - \left[
        \frac{l(l+1)-2}{4}
        \right]^2
    \end{equation}
    for $l\ge 2$ in which case $\chi$ and $\phi$ can be arbitrary linear combinations of spherical harmonics with given $l$ number.\footnote{Notice that we excluded $l=1$ case earlier on and $l=0$ does not contribute anything new.} Notice that when $l=2$, one obtains $\Lambda =0$ and $\Lambda < 0$ whenever $l$ is larger. \\ Thus, we can formulate our main result:
    \begin{remark} \label{thm1}
    All solutions to the  system of equations \eqref{einstein} and \eqref{maxwell} are of the form:
    \begin{align}
    \begin{split}
        \gamma_{ab}^{(1)} &= -2\left(
        \mathring{\epsilon}_{c(a}\mathring D_{b)} \mathring D^c
        \right) \left(\mathring{\Delta} - 2 \right) \chi + \left(
        \mathring{D}_a \mathring{D}_b - \frac{1}{4} q_{ab} (\mathring\Delta -2)
        \right) \Phi +\mathring{D}_a\mathring D_b f\\
        Z_{a} &= -\Psi \epsilon_{ac} \mathring D^c (\mathring\Delta +2) \chi + \frac{1}{2}\Psi \mathring{D}_a f,
    \end{split}
\end{align} where
    \begin{itemize}
        \item $\chi = 0, \Phi = c$ when $\Lambda > 0$
        \item $\chi = \sum_{m=-2}^2 a_{2m} Y_{2m}$,$\Phi = c+ \sum_{m=-2}^2 b_{2m} Y_{2m}$ when $\Lambda = 0$
        \item $\chi = \sum_{m=-l}^l a_{lm} Y_{lm}$,$\Phi = c+ \sum_{m=-l}^l b_{lm} Y_{lm}$ when $4Q^2 \Lambda = 1 - \left[ \frac{l(l+1) -2}{4}\right]^2$ for $l\in \mathbb{N}_{>2}$
        \item $\chi = 0, \Phi = c$ when $\Lambda < 0$ and not included above,
    \end{itemize}
    where $a_{lm}, b_{lm}, c$ are constants
    and $f$ is an arbitrary function which does not encode any physical degrees of freedom. \\ Moreover, a necessary condition for the horizon to be still a marginally outer trapped surface (MOTS) after deformation is \cite{Li:2015wsa}
    \begin{equation}
        \int_{S^2} \gamma^{(1)} > 0
    \end{equation}
    which implies $c>0$.
    \end{remark}
\section{Discussion}\label{discussion}
\subsection{Physical interpretation} \noindent
Stationary spacetime is static when its time-like Killing vector is hypersurface orthogonal, it means
\begin{equation}
    K_{[a} \nabla_b K_{c]} =0.
\end{equation}
In the case of linear deformations on the static background\footnote{In fact, it may be shown that \eqref{general} is the most general static, degenerate Killing horizon in Einstein-Maxwell theory \cite{Kunduri:2008tk}}, it reads (up to the first order in $\epsilon$)
\begin{align}
    \begin{split}
        d h^{(1)} &= 0 \\
        d F^{(1)} &= 0.
    \end{split}
\end{align}
It can be checked that this is equivalent to $\chi = 0$. Thus, when non-trivial deformations exists, we still can have static spacetimes. It would be of interest to find such solutions. For simplicity, one could restrict themselves to the axially symmetric case (it means that $\Phi$ is a linear combination of a constant and of $Y_{l,m=0}$ spherical harmonic). Unfortunately, when $\Lambda \neq 0$ standard techniques of finding static, axially symmetric solutions do not work and so it is highly non-trivial task. On the other hand, it could be reasonably simpler to find them numerically around the horizon. \\
So far we were mainly focused on the non-generic case. Perhaps more interesting case is the generic one in which only the spherically symmetric deformations exist. It suggests (although does not prove!) that  certain spacetimes (e.g. Ernst ones) cannot be generalized to the case of non-vanishing cosmological constant. In particular, a stationary embedding of the electrically charged Reissner-Nordstr\"om black hole into the magnetic field\footnote{Note that generic Reissner-Nordstr\"om-Melvin suffer from the conical singularity and thus are excluded since we are working with smooth objects from the first.} seems to be excluded. Of course, it may happen that such spacetimes exist, but their leading-order deformations are just spherically symmetric. Nevertheless, it is a noticeable difference in contrast to the $\Lambda = 0$ case. \\
In the asymptotically flat setting, one could also have Majumdar-Papapetrou spacetime, which describes multi-centered extremal black holes in which gravity and electromagnetism balance each other. As checked in \cite{Li:2018knr}, their leading-order deformation is exactly spherically symmetric, so our result cannot tell anything about the existence of such black holes\footnote{In fact, multi-centered black hole solutions with $\Lambda > 0$ are known \cite{Kastor:1992nn}. However, they do not possess any Killing vector and thus are excluded from our analysis.}. \\
It seems from this discussion that it is worth investigating is going beyond the first order perturbation in $\epsilon$. The extremely simple form of the solutions in the first order (at least, for generic values of charge) encourages one to go beyond that. If the spherical symmetry persisted, it would be a rather strong argument for the uniqueness of the Reissner-Nordstr\"om-(A)dS spacetimes. We plan to address it in the future work.
\subsection{Connection to the previous work} \noindent
As already mentioned, the systematic study on the extremal horizons' transverse deformations was initiated in \cite{Li:2015wsa} with further developments in \cite{Fontanella:2016lzo, Li:2018knr, Kolanowski:2019wua}. In particular, in \cite{Li:2018knr} all axially symmetric deformations, which consists a 3-dimensional family (including spherically symmetric one), of Reissner--Nordstr\"om were found.\\ We were able to remove this symmetry assumption in our work, discovering 11-dimensional space of allowed deformations. 
Obviously, since Eq. \eqref{einstein} and \eqref{maxwell} are linear, one can superpose solutions which are axially symmetric with respect to the different choices of axis. (Whether they are realized as a solution to the non-linear Einstein-Maxwell equations is an entirely different matter, which lies beyond the scope of this work.) However, they span only 7-dimensional space, and thus they do not cover all solutions presented here. Unfortunately, we are not aware of any extremal black holes solutions without spatial symmetry, so we do not know whether those linearized deformations can be embedded into the non-linear solution. Of course, from the rigidity theorem, it follows that such solutions could not be asymptotically flat. Axially-symmetric case is much better understood and large portion of the solutions space was embedded (into Ernst solutions and very special cases of Kerr-Melvin ones).  \\
The problem of finding solutions to \eqref{einstein} and \eqref{maxwell} with non-zero cosmological constant and charge were not investigated so far in the literature\footnote{Having finished this work, we were informed by James Lucietti about his unpublished result regarding classification of the axially-symmetric deformations of the Reissner-Nordstr\"om-(A)dS black holes which agrees with the solutions presented here.}. Since for generic charge we find only spherically symmetric deformations, there is no problem with an embedding into well-known solutions. Much more interesting are backgrounds with non-generic charges, since they admit a plethora of deformations. Unfortunately, we are not aware of any exact solutions to the Einstein-Maxwell equations in which such particular values of charges were distinguished, so for now their meaning is rather obscure. Even if they exist, they seem to be highly non-generic, even for the standards of the extremal black holes, in vast contrast with the $\Lambda = 0$ case. Perhaps this work could serve as a suggestion that this fine-tuning is needed for constructing new interesting solutions.
\subsection{Generalizations and further work}
\subsubsection{Higher dimensional horizons} \noindent
For now we restricted ourselves only to the four-dimensional spacetimes. Nevertheless, equations for the deformations in the Einstein--Maxwell theory were derived in an arbitrary dimension \cite{Li:2018knr}. It can be easily seen that 
\begin{equation}
    \tilde{\gamma}_{ab} = \gamma_{ab} + \frac{2}{\Psi} D_{(a} Z_{b)}
\end{equation}
is still a gauge invariant quantity\footnote{It would be still true if one omitted symmetrization in $D_a Z_b$. Such non-symmetric $\tilde{\gamma}_{ab}$ may be proven to be more useful.}. Hodge decomposition on $S^d$ reads
\begin{equation}
    Z_a = D^b \omega_{ab} + D_a f,
\end{equation}
where $\omega_{ab}$ is a gauge-invariant two-form (defined up to the term which is $\star d \star$-exact). It seems that one could rewrite all equations using those invariant objects and hopefully solve them again due to the simplicity offered by a spherical symmetry. We hope to address it in the future.  \\
Moreover, in higher-dimensional theories different matter fields could be included (e.g. Chern-Simons term in odd-dimensional ones), those were described in detail in \cite{Fontanella:2016lzo}. Of course, one should start by investigating the simplest possible horizons, it means those which cross-sections are maximally-symmetric. In particular, in $d=5$, all homogenous horizons in the Einstein-Maxwell-Chern-Simons theory were classified in \cite{Kunduri:2013gce}.
\subsubsection{Beyond spherical symmetry} \noindent
Our arguments so far depend heavily upon the assumption of the spherical symmetry of the background. Obviously, the most interesting background would be Kerr-Newman-(A)dS which is only axially symmetric. Unfortunately, we were not able to construct gauge invariant quantities like $\tilde{\gamma}_{ab}$ above. Instead of looking for them, one could split $\gamma_{ab}, Z_a$ into Fourier modes with respect to the Killing vector field $\partial_\phi$. This reduces the problem to the system of (undetermined, due to the gauge symmetry) ODEs\footnote{Indeed, investigation of the axially symmetric deformations of the Kerr-AdS in \cite{Li:2018knr} relies exactly on this}. Unfortunately, solving such a system is again a non-trivial task. In the background of both Kerr and Kerr-AdS it was shown that the only axially symmetric solutions are gauge equivalent to the Kerr and Kerr-AdS itself, respectively. It would be of interest from the point of view of both AdS/CFT correspondence and uniqueness of black holes to determine whether this is still true without additional symmetry assumptions.
\section*{Acknowledgment} \noindent
It is a pleasure to thank Eryk Buk, Jerzy Lewandowski and James Lucietti for useful discussions. This work was financed from budgetary funds for science in 2018-2022 as a research project under the program "Diamentowy Grant".
\begin{appendix}
\section{Technical details}
\label{tech_app} \noindent
In this Section we will describe in more details how Theorem \ref{thm1} can be derived. For simplicity, we will omit subscript $(1)$.  \\
Let us start by a simple observation that a quantity
\begin{equation}
    \tilde{\gamma}_{ab} := \gamma_{ab} - \frac{2}{\Psi} D_{(a} Z_{b)}
\end{equation}
is gauge invariant. Moreover, we have Hodge decomposition of $Z_a$:
\begin{equation}
    Z_a = \Psi \epsilon_{ac}D^c S + \Psi D_a f
\end{equation}
in which $f$ is pure gauge and $S$ is gauge-invariant. (Since $\Psi$ is a non-vanishing constant, it can multiply both terms without loss of generality.) For simplicity, we will multiply both Eq. \eqref{einstein} and \eqref{maxwell} by $r_{+}^2$. In this way we describe everything in terms of geometrical objects associated with $\mathring{q}_{ab}$. Inserting $\tilde{\gamma}_{ab}$ to \eqref{maxwell} leads to
\begin{equation}
    \Psi \mathring{D}^b \left(
    \tilde{\gamma}_{ab} - \frac{3}{2} \tilde{\gamma} \mathring{q}_{ab}
    \right) = -2 \left(
    \mathring\Delta \delta^b_c - \delta^b_c - \mathring{D}_a \mathring{D}^b 
    \right) Z_b,
\end{equation}
where $\tilde{gamma} = q^{ab} \tilde{\gamma}_{ab}$.
Since the left-hand side is gauge-invariant, so must be the right-hand one. Indeed, it is equal to $-2 \Psi\epsilon_{ab} D^b \mathring{\Delta} S$. Any $\tilde{\gamma}_{ab} - \frac{3}{2}\tilde{\gamma} q_{ab}$ satisfying this equation can be uniquely decomposed into the trace-free and divergence-free part, since there are no non-vanishing trace-free, divergence-free symmetric tensors on $S^2$. The most general divergence-free tensor can be written as:
\begin{equation}
    s_{ab} = \left(\mathring D_a \mathring D_b - (\mathring{\Delta} + 1) \mathring{q}_{ab} \right) \Phi. \label{div_free}
\end{equation}
Notice that $\mathring{q}^{ab} s_{ab}$ is an arbitrary function with an exception that it does not possess $l=1$ components, which follows from the following \begin{lemma}
Let $s_{ab} = s_{(ab)}$, $\mathring{D}^a s_{ab} = 0$. Then 
\begin{equation}
    \int_{S^2} Y^\star_{1m} \mathring{q}^{ab} s_{ab} = 0.
\end{equation}
\label{lemma1}
\end{lemma} \noindent
We will prove this lemma by contradiction.
We already constructed solutions with an arbitrary trace without $l=1$ components, so we can assume without loss of generality that $\mathring{q}^{ab} s_{ab} =: b$ has only a dipole part. Then, $D^b f$ is a conformal Killing vector field on $S^2$ and 
\begin{equation}
    \mathring{D}_a \mathring{D}_b f = \frac{1}{2} \mathring{\Delta} f \mathring{q}_{ab}.
\end{equation}
Let us calculate:
\begin{equation}
    \mathring{D}^a \left(
    s_{ab} \mathring{D}^b f
    \right) = s_{ab} \mathring{D}^a \mathring{D}^b f = \frac{1}{2} s \mathring{\Delta} s.
\end{equation}
Integrating this expression on $S^2$ we obtain
\begin{equation}
    0 = \int_{S^2} \mathring{D}^a \left(
    s_{ab} \mathring{D}^b f
    \right) = \int_{S^2} \frac{1}{2} s \mathring{\Delta} s = - \frac{1}{2} \int_{S^2} \mathring{D}^a s \mathring{D}_a s
\end{equation}
Thus we conclude that $s = \textrm{const.} = 0$. This proves this lemma and shows that \eqref{div_free} is the most general divergence-free symmetric $2$-expression. We are thus left with a task of finding any solution to the non-homogenous problem. Natural ansatz is
\begin{equation}
    s_{ab} = 2\mathring{D}_{(a}\epsilon_{b)c} U
\end{equation}
which leads to
\begin{equation}
    \epsilon_{bc} \mathring{D}^c (\mathring{\Delta} + 2) U = - 2 \epsilon_{bc} \mathring{D}^c \Delta S,
\end{equation}
which can be solved by
\begin{align}
    \begin{split}
        U &= 2 \mathring{\Delta} \chi \\
        S &= - (\mathring{\Delta} + 2) \chi
    \end{split}
    \label{max_sol}
\end{align}
unless $S$ has a non-vanishing dipole part. This case can be excluded in the way very similar to the Lemma \ref{lemma1}. Indeed, without loss of generality, we may assume that $S$ has only dipole part. Then, $- 2 \epsilon_{bc} \mathring{D}^c \Delta S =: K_b$ is a Killing covector field and let $\mathring{D}^a s_{ab} = K_b$. One can easily calculate:
\begin{equation}
    \mathring{D}^a \left(s_{ab} K^b \right) = K^2
\end{equation}
Integrating both sides, we get $K^2 = 0$. Thus, the most general solution to \eqref{maxwell} is 
\begin{align}
    \begin{split}
        \tilde{\gamma}_{ab} - \frac{3}{2} \tilde{\gamma} q_{ab}  &= -4\mathring\epsilon_{c(b} \mathring{D}_{a)} \mathring{D}^c  \mathring{\Delta} \chi + \left(
        \mathring{D}_a \mathring{D}_b - \mathring{q}_{ab} (\mathring\Delta +1)
        \right) \Phi \\
        Z_a &= -\Psi \epsilon_{ab} \mathring{D}^b (\mathring{\Delta} + 2) \chi.
    \end{split}
\end{align}
After an easy algebra, one gets \eqref{sol_to_max}. Inserting it to \eqref{einstein} leads to
\begin{equation}
    \left(\mathring{D}_a \mathring{D}_b - \frac{1}{2} q_{ab} \mathring{\Delta} \right) W(\Phi) + \left(
    \mathring{\epsilon}_{c(b} \mathring{D}_{a)} \mathring{D}^c
    \right) V(\chi) = 0, \label{final}
\end{equation}
where
\begin{align}
    W(\Phi) &= \left(-\frac{1}{4}\mathring{\Delta} + \frac{1}{2} - 2 r_+^2 \Psi^2
    \right) \Phi \\
    V(\chi) &= \mathring{\Delta} \left(\mathring{\Delta} -2 + 8r_+^2 \Psi^2 \right) \chi
\end{align}
Taking two divergences of \eqref{final} leads to
\begin{equation}
    0 = \frac{1}{2} \mathring{\Delta} (\mathring{\Delta} + 2) W
\end{equation}
which holds only when $W$ is a linear combination of $l=0$ and $l=1$ spherical harmonics and then
\begin{equation}
    \left(\mathring{D}_a \mathring{D}_b - \frac{1}{2} \mathring{q}_{ab} \mathring{\Delta} \right) W = 0
\end{equation}
is automatically satisfied and one is left with 
\begin{equation}
    \left(\mathring{\epsilon}_{c(b} \mathring{D}_{a)} \mathring{D}^c \right) V = 0.
\end{equation}
Multiplying by $\mathring{\epsilon}_d^{\ b}$ leads to
\begin{equation}
    2\left( \mathring{D}_a \mathring{D}_d  - \frac{1}{2} \mathring{q}_{ad} \mathring{\Delta} \right) V = 0
\end{equation}
and so also $V$ is a linear combination of $l=0$ and $l=1$ spherical harmonics. However, we already deduced that $\chi$ cannot have $l=1$ components and its $l=0$ component was set to zero and thus $V = 0$. Also $l=1$ part of $\Phi$ is meaningless and thus $W = \textrm{const.}$ which leads to \eqref{final_eq}.
\end{appendix}
\bibliographystyle{plain}
\bibliography{bibl}

\begin{thebibliography}{10}

\bibitem{Carter:1971zc}
B.~Carter.
\newblock {Axisymmetric Black Hole Has Only Two Degrees of Freedom}.
\newblock {\em Phys. Rev. Lett.}, 26:331--333, 1971.

\bibitem{chrusciel2012stationary}
Piotr~T Chru{\'s}ciel, Jo{\~a}o~Lopes Costa, and Markus Heusler.
\newblock Stationary black holes: uniqueness and beyond.
\newblock {\em Living Reviews in Relativity}, 15(1):1--73, 2012.

\bibitem{Chrusciel1994}
Piotr~T Chrusciel and Robert~M Wald.
\newblock On the topology of stationary black holes.
\newblock {\em Classical and Quantum Gravity}, 11(12):L147, 1994.

\bibitem{Dobkowski-Rylko:2018nti}
Denis Dobkowski-Rylko, Wojciech Kami\'nski, Jerzy Lewandowski, and Adam
  Szereszewski.
\newblock {The Near Horizon Geometry equation on compact 2-manifolds including
  the general solution for $g>0$}.
\newblock {\em Phys. Lett. B}, 785:381--385, 2018.

\bibitem{Fontanella:2016lzo}
A.~Fontanella and J.~B. Gutowski.
\newblock {Moduli Spaces of Transverse Deformations of Near-Horizon
  Geometries}.
\newblock {\em J. Phys. A}, 50(21):215202, 2017.

\bibitem{hawking1972black}
Stephen~W Hawking.
\newblock Black holes in general relativity.
\newblock {\em Communications in Mathematical Physics}, 25(2):152--166, 1972.

\bibitem{Herdeiro:2016plq}
Carlos A.~R. Herdeiro and Eugen Radu.
\newblock {Static Einstein-Maxwell black holes with no spatial isometries in
  AdS space}.
\newblock {\em Phys. Rev. Lett.}, 117(22):221102, 2016.

\bibitem{PhysRev.164.1776}
Werner Israel.
\newblock Event horizons in static vacuum space-times.
\newblock {\em Phys. Rev.}, 164:1776--1779, Dec 1967.

\bibitem{Kastor:1992nn}
David Kastor and Jennie~H. Traschen.
\newblock {Cosmological multi - black hole solutions}.
\newblock {\em Phys. Rev. D}, 47:5370--5375, 1993.

\bibitem{Kolanowski:2019wua}
Maciej Kolanowski, Jerzy Lewandowski, and Adam Szereszewski.
\newblock {Extremal horizons stationary to the second order: new constraints}.
\newblock {\em Phys. Rev. D}, 100(10):104057, 2019.

\bibitem{Kunduri:2008tk}
Hari~K. Kunduri and James Lucietti.
\newblock {Uniqueness of near-horizon geometries of rotating extremal AdS(4)
  black holes}.
\newblock {\em Class. Quant. Grav.}, 26:055019, 2009.

\bibitem{Kunduri:2013gce}
Hari~K. Kunduri and James Lucietti.
\newblock {Classification of near-horizon geometries of extremal black holes}.
\newblock {\em Living Rev. Rel.}, 16:8, 2013.

\bibitem{Lewandowski:2002ua}
Jerzy Lewandowski and Tomasz Pawlowski.
\newblock {Extremal isolated horizons: A Local uniqueness theorem}.
\newblock {\em Class. Quant. Grav.}, 20:587--606, 2003.

\bibitem{Lewandowski:2004sh}
Jerzy Lewandowski and Tomasz Pawlowski.
\newblock {Quasi-local rotating black holes in higher dimension: Geometry}.
\newblock {\em Class. Quant. Grav.}, 22:1573--1598, 2005.

\bibitem{Li:2015wsa}
Carmen Li and James Lucietti.
\newblock {Transverse deformations of extreme horizons}.
\newblock {\em Class. Quant. Grav.}, 33(7):075015, 2016.

\bibitem{Li:2018knr}
Carmen Li and James Lucietti.
\newblock {Electrovacuum spacetime near an extreme horizon}.
\newblock {\em Adv. Theor. Math. Phys.}, 23:1903--1950, 2019.

\bibitem{Mazur1982}
Pawel~O Mazur.
\newblock Proof of uniqueness of the kerr-newman black hole solution.
\newblock {\em Journal of Physics A: Mathematical and General}, 15(10):3173,
  1982.

\bibitem{PhysRevLett.34.905}
D.~C. Robinson.
\newblock Uniqueness of the kerr black hole.
\newblock {\em Phys. Rev. Lett.}, 34:905--906, Apr 1975.

\end{thebibliography}
\end{document}